# MANAGEMENT OF DANGEROUS GOODS IN CONTAINER TERMINAL WITH MAS MODEL

**Mansoriya Hamidou**[*], **Dominique Fournier**[*], **Eric Sanlaville**[*], **Frédéric Serin**[*]

[*] Litis Laboratory - University of Le Havre, France

([*]mansoriya.hamidou, [*]dominique.fournier, [*]eric.sanlaville, [*]frederic.serin)@univ-lehavre.fr

**ABSTRACT**
In a container terminal, many operations occur within the storage area: containers import, containers export and containers shifting. All these operations require the respect of many rules and even laws in order to guarantee the port safety and to prevent risks, especially when hazardous material is concerned. In this paper, we propose a hybrid architecture, using a Cellular Automaton and a Multi-Agent System to handle the dangerous container storage problem. It is an optimization problem since the aim is to improve the container terminal configuration, that is, the way hazardous containers are dispatched through the terminal to improve its security. In our model, we consider containers as agents, in order to use a Multi-Agent System for the decision aid software, and a Cellular Automaton for modelling the terminal itself. To validate our approach many tests have been performed and the results show the relevance of our model.

## 1. INTRODUCTION

This paper proposes a dynamic technique to manage the storage of containerized dangerous goods in a terminal. This work aims at maintaining the safety of a terminal during all the handling operations that can be executed in such areas.

More precisely, our research is about stacking activities and dangerous containers storage in a port terminal. The problem is: how to position hazardous containers in compliance with physical constraints and regulations? The International Maritime Dangerous Goods (IMDG) Code, available on IMO web site (International Maritime Dangerous Goods 2013), classifies dangerous goods into 9 main classes (Table **1**). Their stockpiling must respect regulation and separation rules for each class. Our aim is to maintain a safe configuration of the terminal. The management of handling equipment is outside the scope of this paper. Methods for the scheduling of Straddle Carrier (SC) missions, and the subsequent routing, are investigated in other papers, see (Lesauvage, Balev and Guinand 2011) and (Balev, et al. 2009).

In the following, we first present more precisely our problem and some related works. Then, our multi-agents hybrid architecture and the behaviours of our agents are presented. Thereafter our tests and results are discussed.

## 2. PROBLEM DESCRIPTION AND LITERATURE

In this section, we will describe the problem of dangerous container storage: first we define the storage area structure we use. Then dangerous goods' classes are determined with examples of security rules. After that, we explain the objective of our work.

### 2.1. Problem Description

#### 2.1.1. Storage Area Structure

A container terminal is a part of a port where containers are stored and handled. The storage area (yard) is divided in blocks. On each block containers are arranged in rows and slots (piles of at most 4 containers high); see Figure **1**.

Spaces between two rows allow the handling equipment circulation. Handling equipment is required for terminal management. It transfers containers within terminal and tranship them. Common types of handling equipment are chassis based transporter, straddle carriers, quay crane, rubber tired gantry crane and rail mounted gantry crane (Stahlbock and Voss 2008).

In a terminal, there are three main activities concerning containers:

- **Unloading:** containers are discharged from a ship or other transport mode like trucks or train, to be transferred to the storage area using handling equipment.
- **Staking:** containers are stored on the area dedicated to them, respecting physical constraints and regulations.
- **Loading:** containers leave storage area and are loaded to be transported on train or ship.

This paper focuses on the stacking activities, and the storage area where containers are moved by Straddle Carriers. When a container is moved from one place to another, within the terminal, we talk about a "movement".

Figure 1 Storage Area Structure

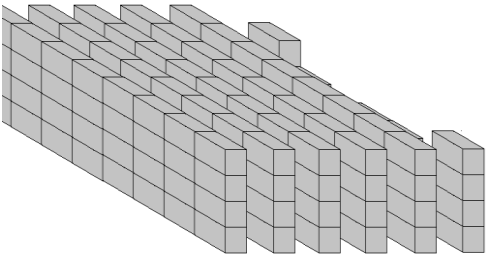

### 2.1.2. Dangerous Containers

Containers are boxes which contain goods. These goods can be dangerous and are then called hazardous materials or *dangerous* goods. This means articles or materials potentially dangerous for people or environment. It includes items of common use, such as aerosol cans, perfumes, and paints (Dangerous goods definition).

The nine IMDG classes of dangerous goods are listed in the following table:

Table 1 IMDG Classes

| Class # | Dangerous Goods |
|---|---|
| 1 | Explosives |
| 2 | Gases |
| 3 | Flammable liquids |
| 4 | Flammable solids; substances liable to spontaneous combustion; substances which, in contact with water, emit flammable gases |
| 5 | Oxidizing substances and organic peroxides |
| 6 | Toxic and infectious substances |
| 7 | Radioactive material |
| 8 | Corrosive substances |
| 9 | Miscellaneous dangerous substances and articles |

Storage constraints exist for each class. The containers of a particular class cannot be stored next to another, or must be separated from them by a fixed distance. An example of separation rules is cited below, see (IDIT 2013).

Flammable liquids containers (class 3) must be separated by:
- Distance F from explosives (class 1). F equals to
  $$F = 4.8 \times Q^{13}$$
  where:
  F is a separation distance in meters; and
  Q is the explosive net weight in kilograms.
- 30 meters from gases (class 2).
- 7 meters from radioactive.
- etc.

### 2.1.3. Objective

To explain the objective of this work, we use an example.

$N$ Containers of different types $T_1$, $T_2$,.. $T_r$ are packed together on one terminal. For simplicity, it shall be supposed in that presentation that the terminal is composed of one unique block of n rows. According to its type and the typology presented before, the well-being of one container can be evaluated. For instance, considering a dangerous container of radioactive type as in previous section, its well-being depends on the number of containers of any dangerous type present in its neighbourhood (defined in terms of Euclidean distance). Generalizing this observation, it is easy to derive a well-being value for each container of the terminal, which can be normalized according to all container types. The total well-being value of a whole terminal configuration can be computed as the worse of the well-being values of all containers it contains (an alternative criterion is the sum of well-beings). This is also called fitness function in section 3.

Consider now some initial terminal configuration, associated with its well-being value. The problem consists in changing the configuration through a sequence of transfers (moves of a container from one place to another) so as to optimize the total well-being by minimizing movements number. This optimization problem is not simple to solve because of the different types of containers, and the special dimension of the problem. It is also clear that the optimal configuration does not depend on the initial configuration, but only on the number of containers of each type. Finding this optimal configuration is a problem of placing objects in a three-dimensional environment, so as to allot each type at best with a minimum number of movements.

### 2.2. Literature

As far as we know, there is no work specially dedicated to the storage of containers with dangerous goods in a terminal, excepted in (Salido, Rodriguez-Molins and Barber 2011). They resolved both allocation berth problem and container stacking problem by a set of Artificial Intelligence based heuristics. In the container stacking problem, the objective was the minimization of number of relocation. In this paper, dangerous containers were considered but the constraint was: two dangerous containers must maintain a minimum security distance, but different existing classes and rules of dangerous containers were not been considered.

However, many research papers use agent-based approach to simulate or solve transport logistics problems (Davidsson, et al. 2005). Some of them study the container terminal management problem using Multi-Agent System and their aims focus on various aspects of terminal planning and management (Rebollo, et al. 2000), (Henesey, Wernstedt et Davidsson 2003) and (Thurston and Hu 2002).

In (Kefi, et al. 2007), a MAS approach was used for storing containers respecting their departure time. The authors use two kinds of agents (Container Agents and Interface Agent) in order to optimize the container storage area on a port terminal; their goal was to reduce the transportation cost within the terminal.

All these works reinforce our idea to use a MAS approach to model the management of a port terminal and to solve our problem. Moreover, (Kefi, et al. 2007) used such architecture to perform container storage optimization which has a spatial aspect like our problem.

Other papers used Operational Research techniques to solve container storage problem in a terminal. (Kim and Hong 2006) proposed two methods for determining the relocation of containers: a branch-and-bound algorithm and a decision rule, but it was limited to only 6 stacks by 5 containers high (5 tiers).

In (Kim and Lee 2006), constraint satisfaction technique was used for space allocation to export containers. The objective was the maximization of the equipment efficiency.

The spatial aspect also appears in works on cellular automata (Wolfram 2002). Cellular automata are in particular used by geographers and economists to model the evolution of a population inside a given space (Schelling 1978). As we shall see later, we use a similar model (see section 3.2.2).

## 3. PROBLEM MODELING

In this section, we define Cellular Automata and show the similarity between the block structure and the Cellular Automata architecture. Then, multi-agent systems are defined and described in the container terminal context. Finally, we detail the adopted strategies and the agents' behaviours.

### 3.1. Hybrid Approach

#### 3.1.1. Cellular Automaton structure

A container terminal is a set of three-dimension cubic cells arranged in rows. These properties inspire us to introduce, by similarity of structure, the notion of 3D Cellular Automaton (CA).

A Cellular Automaton is a complex and dynamic system. It is a collection of cells on a grid. Each cell has a "state" among a finite set of states, and evolves through a number of discrete time steps according to a set of rules based on the states of neighboring cells. The grid can be in any finite number of dimensions (Wolfram 2002). If state updates occur synchronously, we speak about synchronous cellular automata, i.e. the states of every cell in the model are updated together. In contrast, in an asynchronous cellular automaton cells are updated individually and independently, in such a way that the new state of a cell affects the calculation of states in neighbor cells.

Thus, each cell of our cellular automaton corresponds to a container place on terminal. It can be free or occupied. The neighbourhood of each cell depends on the container class and its separation rules, it is defined in terms of Euclidean distance, but a transition function is not simple to be expressed. It will correspond to agents' decisions.

#### 3.1.2. Multi-Agent System Model

A Multi-Agent System is a set of physical or virtual autonomous entities, located on an environment. They can coordinate, communicate, negotiate and interact with each other, using their resources and skills, in order to fulfil common and individual goals (Weiss 1999). Our project aims to avoid a coordination center and consequently to introduce local and neighbourhood consideration to proceed the placing of hazardous items.

As dynamic and complex systems, requiring many decision makers with different objectives, dangerous containers storage problem is suitable for distributed solving techniques. The specification of mobility attached to our agents engaged us to use situated agents in the grid and to precise that elements are not fixed in a definitive cell into the CA. Nevertheless, the agents come in, come out, and move into the CA.

The aim is to satisfy container objectives, that is why container centred model is developed. Consequently, containers are considered as agents and they attempt to reach their goals. Each agent has to be placed in a cell, in which its safety rules are respected. They also contribute to reach the global objective.

Container agents have to execute two processes. The first one is the negotiation phase; the second one is the movement phase. The negotiation phase is composed by the following tasks:

First, each agent computes its well-being. We can restrict the number of partners (containers) interacting in the negotiation phase. Candidates are chosen according to their well-being. The next step consists in finding a destination for elected agent(s); the chosen container can be selected before this step or after to consider the well-being enhancement.

Among the strategies intervening to decide the winner of the negotiation, the moving cost can be considered. It depends on the distance.

After that, the agent selected to be moved will execute the movement process. In this process, container searches a new place better than its current position, and moves using resources.

The "search new place" task can be in the first process or the second process. It depends on the strategy chosen and the agent situation.

This model allows us to test various strategies for dangerous container placement or displacement on a terminal. These strategies depend on processes execution of agents.

To summary our model, first we structure these objects using the CA architecture, secondly, we introduce agent based modelling to add communication protocols and behaviours.

### 3.2. CA and MAS Application: Strategies

To solve the dangerous container storage problem, many strategies are implemented using the Cellular Automata and Multi-Agent System approach.

First, some terms definition, useful for describing strategies, are remained. Then two strategies are detailed

### 3.2.1. Notations
- *Neighbourhood*: The neighbourhood of container agent A is a set of containers. These containers are separated from A by a known distance. For example, a radioactive container R neighbourhood is a set of containers which are located within a radius of 6 meters from the container R.
- *Fitness*: the function fit(A) is used to denote the fitness of the container A. It measures its well-being. It is equal to the number of violated rules of container A.
- *Weighted fitness*: the function $fit_w(A)$ is used to denote the weighted fitness of container A. It is equal to the fitness multiplied by a factor between 0 and 1. This factor depends on the neighbourhood size of the container agent A. The weighted fitness is used to choose candidates agent containers.
- *Configuration*: exact location of each type of container in the block. *An Undesirable configuration* is a configuration in which some security rules are not respected.

### 3.2.2. Schelling Strategy
This strategy was inspired by Schelling's segregation model (Schelling 1978). This model was proposed by Thomas Schelling in 70's. It is a Cellular Automaton used to study racial segregation mechanism inside an urban area. A cell of the automaton is an accommodation (flat or house). Its state is the group of its inhabitant, or empty. The inhabitant decides to leave if the percentage of foreigners (relatively to his group) in its neighbourhood exceeds a given threshold. He then moves to any free accommodation. Under very weak initial conditions and a high tolerant threshold, segregation appears between the different groups of inhabitants.

We use a similar model, where inhabitants are replaced by containers. In this strategy, each container agent interacts with its neighbours, and computes its fitness, then its weighted fitness. Agents with weighted fitness equals to the maximum value of weighted fitness of all agents in the block, move randomly within the terminal. Many agents can be moved in the same time. They choose randomly an empty cell. This strategy is repeated until all security rules are respected, or until the movements becomes too high.

If a container agent is selected to be moved (i.e. his weighted fitness equals to the maximum weighted fitness in the block) and is not on the top of the stack, then this agent asks the agents above to be moved. They also move randomly, from the highest to the lowest.

**Advantages**
- Even if agents move to random places, like Shelling's segregation model, this strategy often find a solution.

**Disadvantages**
- The number of movements is very high.
- Some movements are useless.
- Maximal fitness value of all container agents varies considerably between two configurations (after one strategy run).

### 3.2.3. Cognitive Agent Based Strategy (CABS)
Unlike the previous strategy where container agents are reactive, in this one, agents are cognitive. They anticipate before acting. Only one container moves in the same time.

First, all agents compute their fitness, and their weighted fitness. Then they compare their weighted fitness. The ten agents that have the worst (the ten maximum values) weighted fitness compose "the candidates set". According to strategy steps, this set will be reduced until it contains only one element: the agent to be moved.

Hereafter, step run by candidate agents (agents that are members of "candidate set"):
a) Search place in the block: agents search places; they begin by the nearest empty cells. The best place with the best fitness for each agent is saved in its memory. Each agents stop searching when it finds a place which respects all its security rules.
b) Compute utility value: this value measures fitness improving. It is equal to the difference between the current fitness and the future potential fitness of the agent container. Each candidate agent computes its utility.
c) Reduce candidates set: agents having utility equals to the biggest utility of all candidate agents are retained in the "candidates set". The others are deleted from this set. More over the best utility value must be positive or zero, otherwise the strategy running stops. At this level, candidates set cardinality can be one or more.
d) Candidate agents compare fitness of all neighbour agents, and save on its memory the maximum value called "neighbourhoods' maximum fitness". The agents having the neighbourhoods' maximum fitness value equal to the highest one, are kept in the candidates set, the others are removed.
e) If "candidates set" has more than one element, then one container agent is chosen randomly to be the final candidate.
f) The final candidate moves. If one or more containers are placed just above it, they have to be moved. So, they search the best existing places in the terminal without comparing with the current places, and without taking into account the place chosen by the final candidate (so they can take its chosen place).

Neutral containers (i.e. containers that don't store dangerous goods) have always a fitness equals to zero. When they are ordered to be moved they chose the nearest empty cell.

The program stops if all security rules are respected in the block, or all candidate agents don't find a place that improves or stabilises their fitness.

**Advantages**
- Agents anticipate: before moving, agents search places. Then they compare their fitness improving before they decide which one will be moved.
- A container agent does not move until it improves its fitness or stabilizes it.
- Movement number is strongly reduced comparing with Schelling Strategy

**Disadvantages**
- Anticipation is efficient only if the final candidate container is on top of the stack.
- Risk of getting movement cycles: after *x* movements, the block turns back in a previous configuration and same container(s) move(s) indefinitely.
- The strategy can fall in a local minimum: it can be possible that none of agents container within the candidates set finds a new cell improving or stabilizing its fitness. In this case, the program stops without finding a solution.

## 4. TESTS AND RESULTS

To estimate the efficiency of these strategies, tests were executed. To simplify the model, only five types of containers are considered, with realistic separation rules:
- T1: Highly dangerous containers
- T2: containers storing flammable material
- T3: containers storing oxidizing material
- T4: food containers
- T5: neutral containers

The Table **2** below shows these classes with their separation rules:

Table 2 Separation Rules

| Types | T1 | T2 | T3 | T4 | T5 |
|---|---|---|---|---|---|
| T1 | X | 20m | 20m | VN | X |
| T2 | 20m | X | 6m | VN | X |
| T3 | 20m | 6m | X | VN | X |
| T4 | VN | VN | VN | X | X |

Significance of the table:
- *l* m: it means that there must be a separation of *l* meters between two container types. For example a flammable container (T2) must be stored away from highly dangerous containers (T1) for at least 20 meters and away from oxidizing containers (T3) for at least 6 meters.
- VN is the Von Neumann neighbourhood: a food container cannot be close to a dangerous one.
- X: no constraints.

The solution is implemented with Repast Simphony (Repast Simphony). It is an open source toolkit for agent-based modelling and simulation.

In the following, we present a selection of tests and their results. Remember that, the optimization begins from an initial random configuration, in which many security rules are violated. The purpose is to obtain a final configuration where the number of violated rules is minimum or zero if it is possible, with a minimum of movements. In this first phase, the dynamic of containers terminal is not taken into account: during simulation, there is no arrival or departure to or from the terminal.

### 4.1. Schelling vs. CABS

To compare the two strategies defined in section 3.2, we consider one block with the five previously defined container types. The block is composed of 10 rows; each row is composed of 10 container stacks, and each stack is at most four containers high. So, the block contains 400 cells.

75% of cells are occupied by containers. The average percentage of dangerous containers is about 10% of the global traffic. In our tests, the percentage is increased to 15% to verify the robustness of our approach. Food containers (T4) represent 20% and the rest (65%) are neutral containers (T5). Hereafter a recapitulative of the block properties:

Dimensions : 10x10x4  
Filling     : 75%  
%T1         : 1%  
%T2         : 7%  
%T3         : 7%  
%T4         : 20%  

Tests were done on 1000 instances. Results are expressed on the table above:

Table 3 Schelling vs. CABS

| Strategies | | Schelling | CABS |
|---|---|---|---|
| Number of movements | Min | 277 | 62 |
| | Max | 8906 | 130 |
| | Avg | 703,43 | 92,61 |
| % success | | 95,40 % | 99,70% |

**Min**, **Max** and **Avg**, are respectively the minimum, maximum and the average of the number of movements, in case the optimal solution was found.
**% success** is the percentage of optimal solutions found by the strategy among solutions.

During tests, the program is stopped if the solution is found or if the number of movements reaches 10000 movements for Schelling strategy and 1000 movements for CABS.

Among 1000 CABS runs, the program stopped without finding a solution in 3 cases. In these 3 cases the program reached the limit number of movements

(1000 movements), and movement cycles were observed.

To avoid cycles in CABS, some modifications were done: when an agent container moves, it keeps temporarily in memory the last places occupied. When a container searches for a new place, it checks its list of memorized places, and avoids them.

### 4.2. Block Filling Variation

In these tests of CAB Strategy, we try to vary the parameter of the block filling (50%, 70% and 90%), and observe how the strategy is efficient.

Figure 2 Number of Movements When The Block Filling Varies

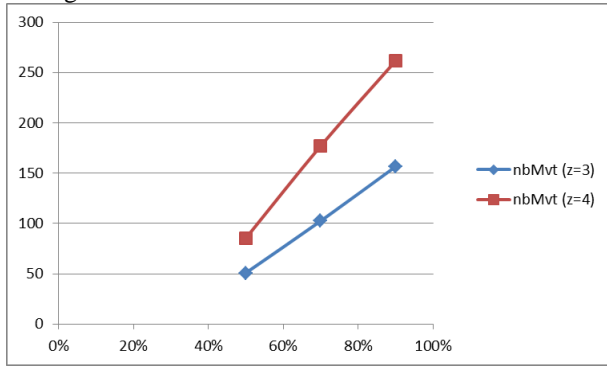

The rate success in the six cases is 100%.
We observe that when the filling percentage increases to high values, the number of movements increases no more than linearly.

### 4.3. Dimension Block Variation

Now we vary the block dimension: the number of rows (X) and the stack high (Z). The other parameters remain unchanged and are kept as before.

Table 4 Test Results: Block Dimensions Variation

|   | X=10 | | X=20 | | X=40 | |
|---|---|---|---|---|---|---|
| Z | *nbMvt* | *Success* | *nbMvt* | *Success* | *nbMvt* | *Success* |
| 2 | 23,87 | 99,9 % | 50,56 | 100 % | 103,35 | 100 % |
| 3 | 51,27 | 100 % | 103,24 | 100 % | 213,56 | 100 % |
| 4 | 91,49 | 99,8 % | 177,07 | 100 % | 355,74 | 100 % |

nbMvt: is the average number of movements.
Time: is the average time of run in seconds.

These tests show that the number of movements is linearly proportional to the number of containers.

### 4.4. T1 Percentage Variation

Here the percentage of T1 containers is allowed to vary. The block dimensions are 10x10x3 and the percentages of T2, T3 and T4 remain unchanged (7%, 7% and 20% respectively).

The results show that from 0 to 3% (6 containers) of T1 containers, almost always an optimal solution is found, and the number of movements increases.
From 4% to 11%, the percentage of success decreases rapidly, from 91% to 1%, until no solution is found for 12%, which corresponds to 27 T1 containers. Note that the number of movements is then non increasing.

Figure 3 The Percentage of Success When T1 Containers Varies

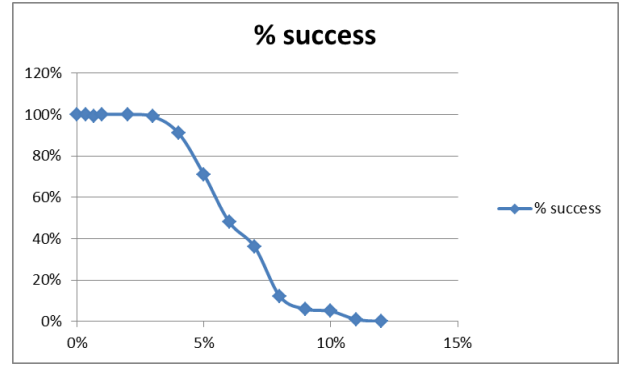

Figure 4 The Number of Movements When T1 Containers Varies

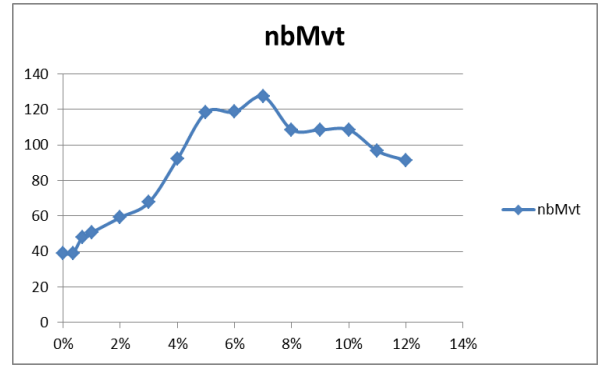

### 4.5. Tests Conclusion

Our decentralized multi-agents model is an efficient way to solve the problem of dangerous container storage when the parameters (percentage of dangerous containers) are realistic. It is efficient even if the initial configuration is created randomly, which is making things more difficult.

In a second step, the model will include the dynamics of the terminal.

## 5. CONCLUSION

The presented framework permits to study different situations. We have tested our container centred approach modelling. The obtained results permit to validate the model and the simulation compared to the ground truth. Complementary test are being realized to take into account the dynamic arrivals and departures of containers.

After validating the behaviour of our intelligent agents, we have introduced some extreme values considering the density of dangerous containers. The results show the existence of a threshold; beyond this threshold our model has difficulties to reach an optimal solution. Nevertheless, these unsuccessful configurations are quite unrealistic.

Now, our model invites us to test specific agent behaviours using the centred container approach. This consists in modifying each intelligent agent to introduce different perceptions of the environment and to change the rules of positioning and fitness computing. So, this modelling permits the practical modularity and flexibility, allowing to test and to adapt strategies to the dynamic context of a container terminal.

## ACKNOWLEDGEMENTS

This work has been supported by the European Community, the Haute Normandie Region and the French State through the "Passage Portuaire" project.

## REFERENCES

Balev, Stephan, Frédéric Guinand, Gaetan Lesauvage, and Damien Olivier. "Dynamic Handling of Straddle Carriers Activities on a Container Terminal in Uncertain Environment - A Swarm Intelligence Approach." *3rd International Conference on Complex Systems and Applications.* 2009.

*Dangerous goods definition.* http://www.businessdictionary.com/definition/dangerous-goods.html (accessed 04 13, 2013).

Davidsson, P, L Henesey, L Ramstedt, J Tornquist, and F Wernstedt. "An analysis of agent-based approaches to transport logistics." *Transportation Research Part C: Emerging Technologies* 13, no. 4 (2005): 255-271. Edited by Elsevier Science Publishers Ltd.

Henesey, L, F Wernstedt, et P Davidsson. «Market-Driven Control in Container Terminal Management.» *Proc. of the 2nd Int. Conference on Computer*, 2003: 377-386.

IDIT. *The Institute of International Transport Law.* http://www.idit.fr (accessed 04 13, 2013).

*International Maritime Dangerous Goods.* 2013. http://www.imo.org/blast/mainframe.asp?topic_id=158 (accessed 04 13, 2013).

Kefi, Meriam, Korbaa Ouajdi, Khaled Guedira, and Yim Pascal. "Container handling using multi-agent architecture." *Agent and Multi-Agent Systems: Technologies and Applications*, 4496 (2007): 685-693. Edited by Springer.

Kim, Kap Hwan, and Gyu-Pyo Hong. "A heuristic rule for relocating blocks." *Computers Operations Research* 33, no. 4 (2006): 940 - 954.

Kim, Kap Hwan, and Jong-Sool Lee. "Satisfying Constraints for Locating Export Containers in Port Container Terminals." *Computational Science and Its Applications - ICCSA 2006* 3982 (2006): 564-573. Edited by Springer Berlin / Heidelberg.

Lesauvage, Gaëtan, Stephan Balev, and Frédéric Guinand. "D²CTS: A Dynamic and Distributed Container Terminal Simulator." *The 14th International Conference on Harbor, Maritime & Multimodal Logistics Modelling and Simulation.* 2011.

Rebollo, M, V Julian, C Carrascosa, and V Botti. "A multi-agent system for the automation of a port container terminal." *Workshop in Agents in Industry. Barcelona*, 2000.

*Repast Symphony.* http://repast.sourceforge.net/ (accessed 04 13, 2013)

Salido, Miguel A., Mario Rodriguez-Molins, and Federico Barber. "Integrated intelligent techniques for remarshaling and berthing in maritime terminals." *Advanced Engineering Informatics* 25, no. 3 (2011): 435-451.

Schelling, Thomas. *Micromotives and macrobehavior.* Norton, Toronto, 1978.

Stahlbock, Robert, and Stefan Voss. "Operations research at container terminals: a literature update." *OR Spectrum* 30, no. 1 (2008): 1-52.

Thurston, Tom, and Huosheng Hu. "Distributed Agent Architecture for Port Automation." *Computer Software and Applications Conference, COMPSAC 2002.* Oxford, England, 2002. 81-87. Edited by IEEE Computer Society.

Weiss, Gerhard. *Multiagent systems: a modern approach to distributed artificial intelligence.* (1999), The MIT Press.

Wolfram, Stephan. *A new kind of science.* Champaign IL USA: Wolfram Media, 2002.